\documentclass[useAMS,usenatbib]{mn2e}

\DeclareSymbolFont{cmletters}{OML}{cmm}{m}{it}
\DeclareMathSymbol{v}{\mathalpha}{cmletters}{"76}

\usepackage{tabularx,ragged2e,booktabs,caption}
\usepackage{amsmath}
\usepackage{amssymb}
\usepackage{epsfig}
\usepackage{graphicx}
\usepackage{ifthen}
\usepackage{latexsym}
\usepackage{rotating}
\usepackage{subfigure}
\usepackage{times,epsf}
\usepackage{txfonts}
\usepackage{varioref}
\usepackage{verbatim}
\usepackage{array}
\usepackage{url}
\usepackage{color}
\usepackage[dvipsnames]{xcolor}
\usepackage[T1]{fontenc}

\newcommand{\be}{\begin{equation}}
\newcommand{\ee}{\end{equation}}
\newcommand{\bea}{\begin{eqnarray}}
\newcommand{\eea}{\end{eqnarray}}

\newcommand\actaa{Acta Astronomica}

\newcommand\apj{Astrophysical Journal}
\newcommand\apjl{Astrophysical Journal Letters}

\newcommand\aap{Astronomy \& Astrophysics}

\newcommand\nat{Nature}
\newcommand\prd{Physical Review D}
\newcommand\mnras{Monthly Notices of the Royal Astronomical Society}

\newcommand\pasj{Publications of the Astronomical Society of Japan}

\newcommand\ARAA{Ann. Rev. Astron. Asrophys.}
\newcommand\jqsrt{Journal of Quantitative Spectroscopy and Radiative Transfer}
\newcommand\araa{\ARAA}

\newcommand{\koral}{\texttt{KORAL}}
\newcommand{\Medd}{\dot M_{\rm Edd}}

\title[Radiative jets]{Powerful radiative jets in super-critical
  accretion disks around non-spinning black holes}
\author[A. S\k{a}dowski, R. Narayan]
       {Aleksander S\k{a}dowski$^1$\footnotemark[1] and Ramesh Narayan$^{2}$\thanks{E-mail: asadowsk@mit.edu (AS); 
rnarayan@cfa.harvard.edu (RN);} \\
        $^1$ MIT Kavli Institute for Astrophysics and Space Research,
77 Massachusetts Ave, Cambridge, MA 02139, USA\\
$^2$ Harvard-Smithsonian Center for Astrophysics, 60 Garden St., Cambridge, MA 02134, USA}

\begin{document}

\maketitle

\label{firstpage}

\begin{abstract}
We describe a set of simulations of super-critical accretion onto a
non-rotating supermassive BH.
The accretion flow is radiation pressure dominated and takes the form
of a geometrically thick disk with twin low-density funnels around the
rotation axis. For accretion rates $\gtrsim 10\Medd$, there is
sufficient gas in the funnel to make this region optically
thick. Radiation from the disk first flows into the funnel, after
which it accelerates the optically thick funnel gas along the axis.
The resulting jet is baryon-loaded and has a terminal density-weighted
velocity $\approx 0.3c$. Much of the radiative luminosity is converted
into kinetic energy by the time the escaping gas becomes optically
thin. These jets are not powered by black
hole rotation or magnetic driving, but purely by radiation.  Their
characteristic beaming angle is $\sim0.2$ radians. For an observer
viewing down the axis, the isotropic equivalent luminosity of total
energy is as much as $10^{48}\,\rm erg\,s^{-1}$ for a $10^7 M_\odot$
BH accreting at $10^3$ Eddington.  Therefore, energetically, the
simulated jets are consistent with observations of the most powerful tidal
disruption events, e.g., Swift J1644. The jet velocity is, however, too low to
match the Lorentz factor $\gamma > 2$ inferred in J1644. There is no
such conflict in the case of other tidal disruption events. Since
favorably oriented observers see isotropic equivalent luminosities
that are highly super-Eddington, the simulated models can explain
observations of ultra-luminous X-ray sources, at least in terms of
luminosity and energetics, without requiring intermediate mass black
holes. The spectrum remains to be worked out.  Finally, since the
simulated jets are baryon-loaded and have mildly relativistic
velocities, they match well the jets observed in SS433. The latter
are, however, more collimated than the simulated jets. This suggests
that, even if magnetic fields are not important for acceleration, they
may perhaps still play a role in confining the jet.

\end{abstract}

\begin{keywords}
  accretion, accretion discs -- black hole physics -- relativistic
  processes -- methods: numerical -- galaxies: jets --- X-rays:
  individual: Sw J1644+57 --- X-rays: individual: SS433
--- X-rays: ULX
\end{keywords}

\section{Introduction}
\label{introduction}

Accretion flows around compact objects are commonly associated with
relativistic jets.  Such jets in active galactic nuclei, X-ray
binaries in the low/hard state and gamma-ray bursts can be extremely
powerful and can reach high Lorentz factors. It is believed that jets
are driven by the rotational energy of a spinning black hole (BH)
extracted via magnetic fields \citep[][BZ]{bz}. The efficiency of
energy extraction depends on the BH spin and the magnetic flux
threading the BH horizon. The maximal power for a given spin is
reached when the magnetic flux saturates at the value characteristic
of the magnetically arrested disk (MAD) state
\citep{igu+03,narayan+03,tchekh11,mtb12} , i.e., when the magnetic
pressure near the horizon on average balances the ram pressure of the
accretion flow. For rapidly rotating BHs, the power of relativistic
jets in the MAD state can exceed $\dot M c^2$
\citep{tchekh11,tchekh+12}, i.e., more energy can go into the jet than
the entire rest mass energy accreted by the BH.  Such relativistic and
magnetic jets have been extensively studied in recent years, both
numerically \citep[e.g.,][]{tchekh10a,tchekh11,tchekh+12} and
analytically \citep[e.g.,][]{penna+bz,lasota+bz}.  Although these
studies have focused predominantly on hot, optically thin,
geometrically thick accretion flows, recent numerical simulations
\citep{sadowski+koral2, mckinney+harmrad} suggest that BH
rotation-driven jets may be produced with comparable efficiency even
in the case of optically thick accretion flows.

Relativistic jets are often invoked to explain the observed X-ray
luminosities of tidal disruption events (TDEs). When a star is
disrupted by a supermassive BH, part of the debris returns to the BH
and is accreted at a very high, significantly super-Eddington
(``super-critical'') mass accretion rate. The isotropic equivalent
X-ray emission can reach as much as $10^{48}\,\rm erg\,s^{-1}$
\citep{bloom11,burrows11}. It is believed that this radiation is
produced in internal shocks within a relativistic jet. However, as
mentioned above, the most powerful jets require accumulation of a
significant amount of magnetic flux, and the disrupted star is
incapable of providing this flux \citep{tchekh+14}. In addition, a
rapidly spinning BH is required to produce an effective
Blandford-Znajek (BZ) jet.

The initial mass accretion rate in a TDE is super-Eddington, which is
very different from the highly sub-Eddington, radiatively inefficient
regime that is usually considered when studying magnetic jets. A
super-Eddington accretion disk is dominated by radiation pressure and
the gas is optically thick. It was predicted already by
\cite{paczynskiwiita-80} that such disks will be super-luminous. What
is more, the large thickness of the disk naturally collimates
radiation along the rotation axis, producing a significantly
super-Eddington flux of energy in the funnel
\citep{sikora-81,narayan+83} which can accelerate gas to
mildly relativistic velocities \citep{sikorawilson-81}. Such
radiatively driven jets should occur in any super-critical accretion
flow and should be present even if the BH is non-rotating or if there
is little magnetic flux threading the BH horizon.  In this work we
study in detail the properties of such purely radiative (non-BZ) jets
by means of general relativistic (GR) radiation MHD simulations.

The paper is organized as follows. In Section~\ref{s.method} we
describe the numerical method. In Section~\ref{s.results} we present
the results, and discuss in detail the location of the photosphere
(Section~\ref{s.photo}), the properties of the radiative jet
(\ref{s.jet}), the corresponding luminosities (\ref{s.luminosities}),
and the beaming factor (\ref{s.beaming}). In
Section~\ref{s.discussion} we discuss applications to TDEs,
ultraluminous X-ray sources, and SS433.

\section{Numerical methods}
\label{s.method}

We performed a set of three simulations of super-Eddington accretion
on a non-rotating, $a_* = 0$, moderately-supermassive, $M_{\rm
  BH}=3\times 10^5 M_\odot$, black hole. We used the general
relativisic (GR) radiation magnetohydrodynamical (RMHD) code \koral\,
\citep{sadowski+koral,sadowski+koral2}, which evolves gas and magnetic
field in the ideal MHD approximation, together with the radiation
field under the M1 closure scheme \citep{levermore84}. The models were initialized as
optically thick equilibrium tori of gas and radiation in local thermal
equilibrium, threaded by multiple loops of weak magnetic field. The
only difference in the intitial states of the three simulations was
the torus entropy parameter $\cal{K}$ \citep[see][]{penna-limotorus},
which controls the initial torus density and thereby the mass
accretion rate.

The simulations were axisymmetric and were run with a resolution of
$304\times192$ cells in radius and polar angle, with polar cells
concentrated towards the equatorial plane.  Each simulations was
evolved for an exceptionally long time --- roughly 2,000 orbital times
at the innermost stable circular orbit (ISCO). This was possible
because the poloidal magnetic field was amplified and maintained at a
physically reasonable level by means of a mean-field dynamo, as
described in \cite{sadowski+dynamo}. (Without the dynamo, the poloidal
magnetic field would quickly decay because of axisymmetry of these 2D
simulations. On the other hand, without restricting ourselves to 2D,
it would be virtually impossible to run the simulations for such long
durations.) The duration of the simulations was sufficiently long for
the jet to reach equlibrium at the outer boundary of the grid, located
at $r=10000$\footnote{Throughout the paper we adopt $GM/c^2$ as the
  unit of length, and $GM/c^3$ as the unit of time.}.   We used the following formulae for the
absorption and scattering opacities, $\kappa_{\rm abs}$ and
$\kappa_{\rm es}$, respectively,
\bea
\kappa_{\rm abs}&=&6.4\times 10^{22} \rho T^{-7/2}\,\rm cm^2/g,\\
\kappa_{\rm es}&=&0.34 \,\rm cm^2/g.
\eea The model
parameters of the three simulations are given in Table~\ref{t.models}
and are defined as in \cite{sadowski+dynamo}.

The average accretion rates for models \texttt{A}, \texttt{B} and
\texttt{C} were $45$, $310$ and $4800 \Medd$, respectively, where the
critical accretion rate $\Medd$ is defined as, \be
\label{e.medd}
\Medd = \frac{L_{\rm Edd}}{\eta_0 c^2} = 2.44\times10^{18}
\frac{M}{M_{\odot}}{\rm g\,s^{-1}}, \ee where $L_{\rm Edd}=1.25\times
10^{38}M/M_{\odot}\, \rm erg\,s^{-1}$ is the Eddington luminosity, and
$\eta_0$ is the radiative efficiency of a thin disk for zero BH spin:
$\eta_0=0.057$. For the chosen BH mass ($M_{\rm BH}=3\times 10^5
M_{\odot}$), $L_{\rm Edd}=3.75\times 10^{43}\, \rm erg\,s^{-1}$ and
$\Medd=7.32\times 10^{23}\,{\rm g\,s^{-1}}$.

Given the long durations of the runs (Table~\ref{t.models}), the
simulations reached inflow/outflow equilibrium\footnote{Defined as a
  region where the local average radial velocity $v^r$ satisfies
  $|v^r|>r/2t$ for a simulation of duration
  $t$.}  up to radius $r_{\rm eq}\sim 50 $ in the bulk of the disk,
where the gas moves inward on average with a relatively low
viscosity-driven velocity. In the region near the axis, where the gas
flows out with quasi-relativistic velocities (as will be discussed
below), the converged region extends out to the outer boundary at
$r=10000$.

\begin{table}
\begin{center}
\caption{Model parameters}
\label{t.models}
\begin{tabular}{llcccc}
\hline
\hline
Name &\hspace{.5cm}  &$\cal{K}$ &  $t_{\rm max} / (GM/c^3)$ 
& $\langle \dot M \rangle /\Medd$ & $\eta$\\
\hline
\hline
\texttt{A} && $10.0$ & 200,000  &  45 & 5.3\%\\ 
\texttt{B} && $5.0$ & 170,000  &  310 & 4.0\%\\ 
\texttt{C} && $1.0$ & 190,000  &  4800 & 4.6\%\\ 
\hline
\hline
\end{tabular}
\end{center}
Other parameters: BH mass: $M_{\rm BH}=3\times 10^5 M_{\odot}$, BH spin: $a_*=0.0$,
resolution: 304x192, radial extent of the domain: $R_{\rm min}=1.85$ to $R_{\rm max}=10000$,
grid parameters \citep[defined in][]{sadowski+dynamo} $R_0=1.0$,
$H_0=0.6$, 
minimal initial ratio of magnetic to total pressures:
$\beta_{\rm max}=10.0$.
\end{table}

\section{Results}
\label{s.results}

\subsection{Photosphere}
\label{s.photo}

\begin{figure}
  \hspace{.2cm}
 \includegraphics[width=.85\columnwidth]{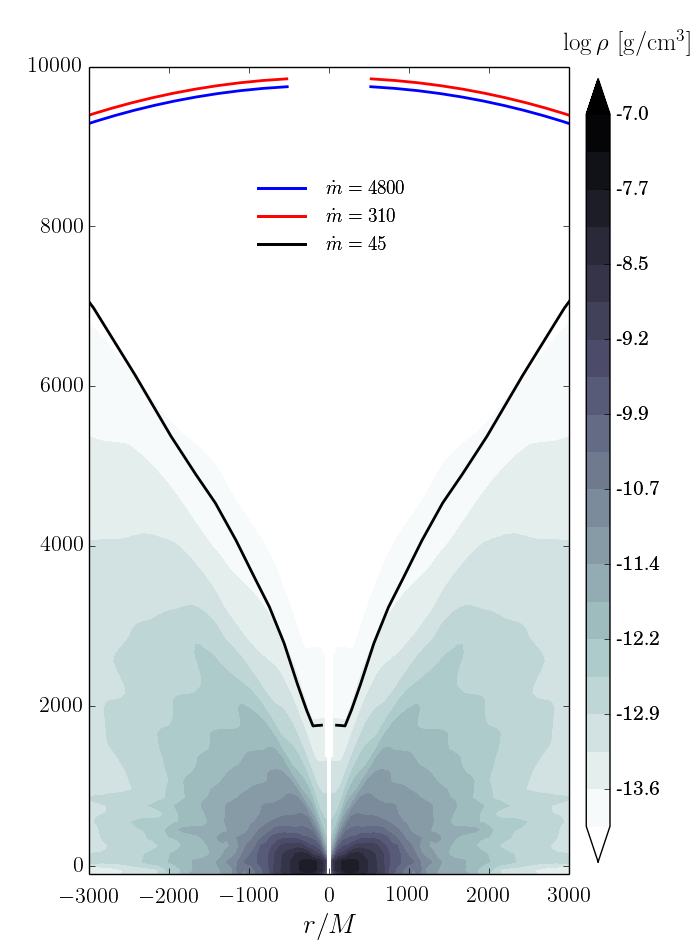}
  \caption{ Photosphere profiles for models \texttt{A} (black),
    \texttt{B} (red) and \texttt{C} (blue lines). Only model
    \texttt{A} has its photosphere inside the computational
    domain. The other two lines represent only the formal location of
    the photosphere which coincides with the domain boundary at
    $r=10000$. The color levels represent the logarithm of density for
    model \texttt{A}.  }
 \label{f.photo}
\end{figure}

The simulated accretion flows are characterized by high,
super-critical accretion rates and form geometrically thick disks (as
expected). Figure~\ref{f.photo} shows the time-averaged gas
distribution in Model \texttt{A}, with color intensity indicating the
mean density. Within the inflow/outflow equilibrium radius ($r_{\rm
  eq}\sim 50$) the disk density scale height corresponds to $h/r\sim
0.3$, a typical value for super-critical and radiatively inefficient
accretion flows.  Although most of the gas is concentrated near the
equatorial plane, a significant amount is blown away to form the wind
region visible in Fig.~\ref{f.photo} at intermediate polar angles. The
lowest gas density is found near the polar axis.  For Model \texttt{A}
(accreting at the lowest rate, $\dot M=45\dot{M}_{\rm Edd}$), the
density on the axis at radius $r=1000$ equals roughly $10^{-13}\rm
g\,cm^{-3}$. This corresponds to an optical depth against scattering
greater than unity over the distance $d=1000\approx 10^{13}\rm
cm$. Thus, a significant fraction of the polar region is optically
thick.

We estimate the radius of the photosphere in the funnel as a function
of polar angle $\theta$ by integrating the optical depth from the
outer boundary of the simulation box down towards the BH along fixed
$\theta$. In computing the optical depth, we allow for relativistic
effects and follow the formalism given in \cite{sadowski+dynamo}
(their eqs. 47-49). The black contours in Fig.~\ref{f.photo} show the
location of the photosphere for model A.  Even for this moderately
super-critical model, the photosphere at the axis is located as far
out as $r=2000$. For larger accretion rates the photosphere moves even
farther out. In fact, for Models \texttt{B} and \texttt{C}, it lies
otuside the computational domain (red and blue lines in
Fig.~\ref{f.photo} show only the formal photospheres coinciding with
the outer edge of the domain).

\subsection{Radiative jet}
\label{s.jet}

\begin{figure}
\hspace{-.75cm}
 \includegraphics[width=.95\columnwidth]{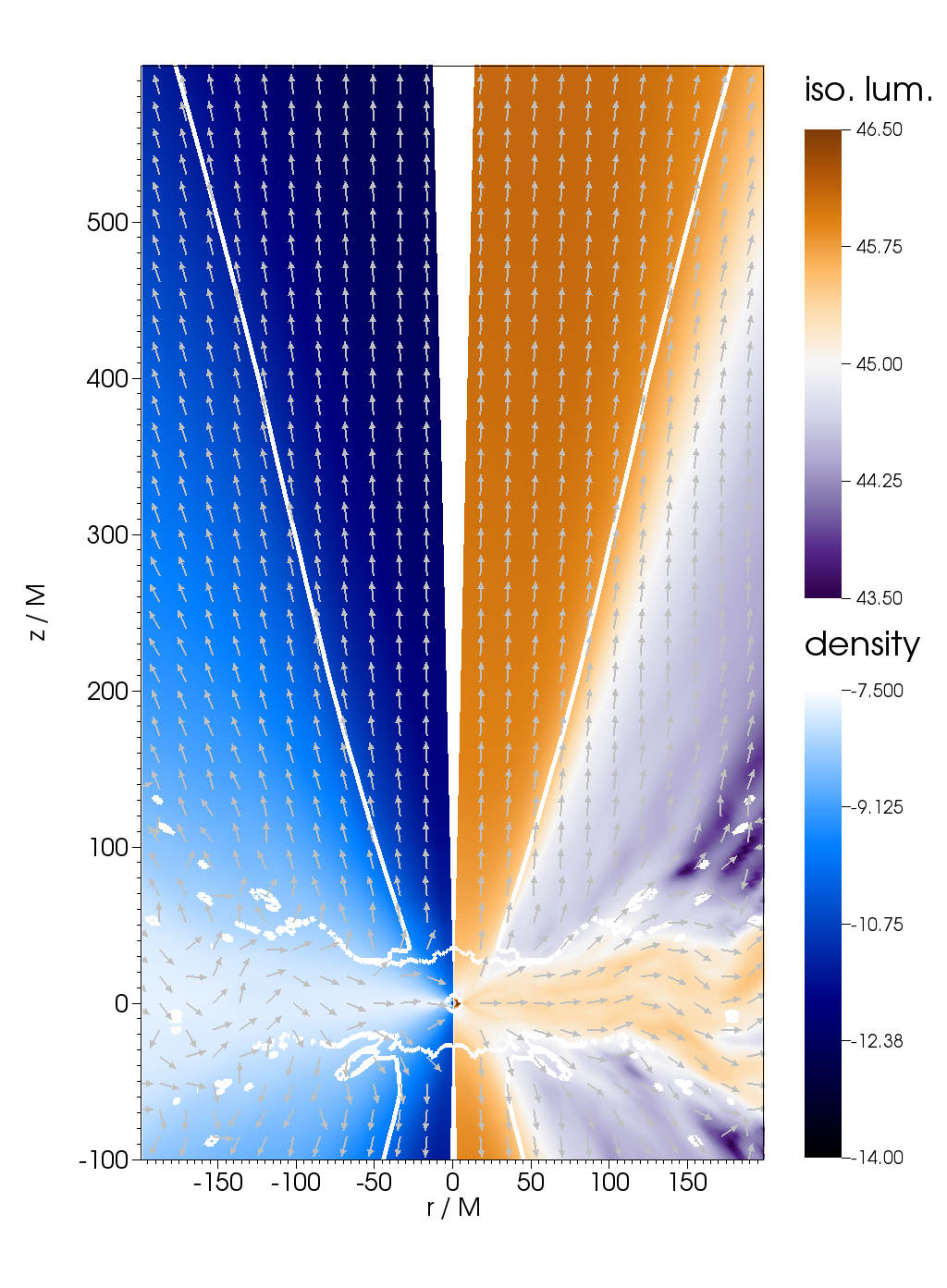}
  \caption{Time-averaged results for model \texttt{A} in the poloidal
    plane. Left Panel: Color shows the distribution of the logarithm
    of density (in $\rm g\,cm^{-3}$) and arrows show the gas
    velocity. The white contour shows the formal border of the jet
    region. Right Panel: Color shows the logarithm of isotropic
    equivalent luminosity of total energy (in $\rm erg\,s^{-1}$, for
    definition see Eq.~\ref{e.isolum}) and arrows show the direction
    of the energy flux. The white contour shows the border.  }
 \label{f.fluxden}
\end{figure}

The left panel of Fig.~\ref{f.fluxden} shows again the density
distribution in Model \texttt{A}, but limited now to the inner regions
and with gas velocity vectors superimposed.  The right panel shows the
magnitude and direction of the energy flux. To explain the
quantity that is plotted, we start with the stress-energy tensors,
$T^\mu_\nu$, $R^\mu_\nu$, of the magnetized gas and radiation, and
note that $-(T^r_t+R^r_t)$ at any point is the radial flux of total
energy (gas, magnetic and radiation) at that point as measured by an
observer at infinity (lab frame). Also, $\rho u^r$, where $\rho$ is
the gas density in the fluid frame and $u^r$ is the fluid
four-velocity in the lab frame, is the radial flux of rest mass
energy. Thus, 
\be
\label{e.enflux}
F^r = -(T^r_t + R^r_t + \rho u^r)
\ee
describes the flux of
``useful energy'' (i.e., rest mass energy subtracted out) as measured
at infinity. Multiplying $F^r$ by $4\pi r^2$ then gives the equivalent
isotropic luminosity ($\rm erg\,s^{-1}$) that one infers from the
measured flux at the given location, \be L_{\rm iso}=4\pi r^2 F^r= -4\pi
r^2 (T^r_t + R^r_t + \rho u^r).
\label{e.isolum}
\ee The colors in the right panel of Fig.~\ref{f.fluxden} show
this quantity, $L_{\rm iso}$, as a function of position, with arrows
indicating the direction of the energy flux in the poloidal
plane\footnote{The direction in the poloidal plane is determined by
  the vector $(\hat F^r,\,\hat F^\theta)$ built from ortonormal radial
  and polar
  components of the energy flux as defined in Eq.~\ref{e.enflux}.}. Over the entire volume, energy
flows out, i.e., $L_{\rm iso} > 0$. In the region of the disk, bound
gas is accreted, while in the wind and jet regions, unbound gas is
ejected. Both correspond to positive $L_{\rm iso}$.  The highest
energy flux is in the region near the polar axis. Interestingly, even
for the relatively low accretion rate of $45\Medd$ in model \texttt{A},
the isotropic luminosity reaches $5\times 10^{45}\, \rm erg/s\approx
130 L_{\rm Edd}$.

To define the region corresponding to the jet, we follow
\cite{sadowski+outflows} and require the local Bernoulli parameter
$\mu$, defined as
\be
\mu=\frac{F^r}{\rho u^r}=-\frac{T^r_t+R^r_t+\rho u^r}{\rho u^r}, 
\label{e.mu}
\ee 
to exceed $0.05$, which
corresponds to the velocity at infinity exceeding roughly $30\%$ of
the speed of light. This definition is not fundamental for our
analysis, but is useful when talking about averaged or integrated
quantities in the funnel.

The white contour in Fig.~\ref{f.fluxden} denotes the border of the
jet region defined according to the criterion given above. It agrees
well with the region of lowest density (shown in the left panel) as
well as the region of highest energy flux (right panel).

\begin{figure}
 \includegraphics[width=.95\columnwidth]{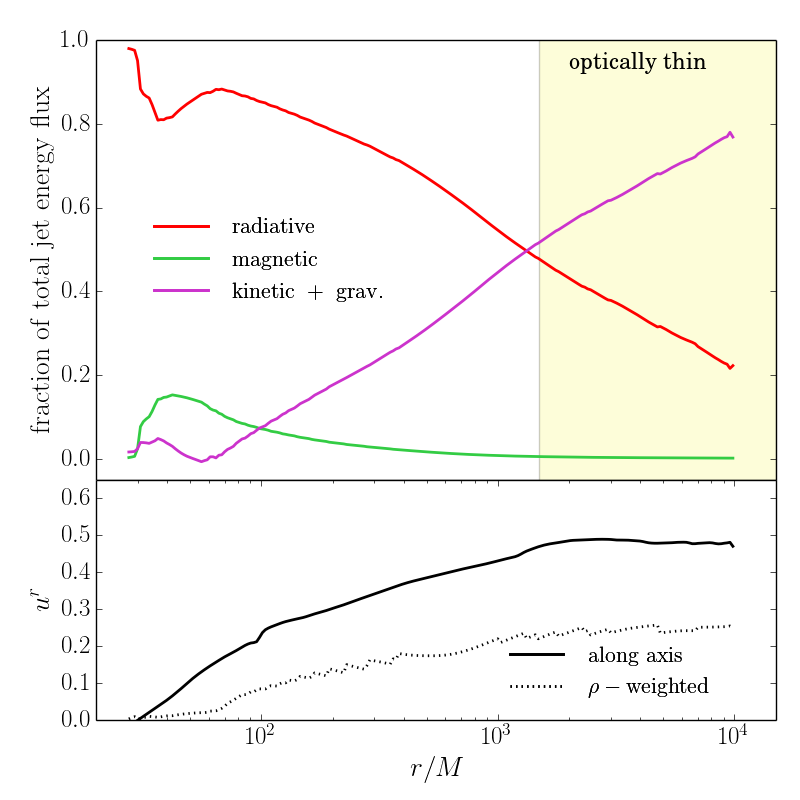}
  \caption{Top panel: Fractional contribution in model \texttt{A} of
    radiative ($R^r_t$), magnetic ($b^2 u^r u_t -b^r b_t$) and kinetic
    plus gravitational ($\rho u^r u_t + \rho u^r$) components of the
    energy flux to the total flux in the jet region as a function of
    distance. The shaded zone corresponds to the optically thin region
    above the photosphere.  Bottom panel: Radial profiles of gas
    velocity in the jet region in model \texttt{A}. The solid line
    shows the gas velocity in the cell nearest to the axis, while the
    dotted line shows the density-weighted average velocity in the jet
    region.  }
 \label{f.enfluxes}
\end{figure}

The top panel of Fig.~\ref{f.enfluxes} shows radial profiles of the
fractional contributions of various components of the energy flux
integrated over solid angle within the jet region.  Red, purple and
green lines correspond to the fluxes of radiative, kinetic, and
magnetic energies, respectively. The flux of thermal energy is
negligible and is not shown.

In the region of the jet closest to the BH ($r\lesssim 1000$), the
energy flux is dominated by radiation which comes from the innermost
and hottest part of the disk.  However, the relative contribution of
the radiative flux decreases with increasing distance from the BH as
radiative energy is converted into kinetic energy (radiation pressure
pushes the gas out). The magnetic contribution is unimportant at all
radii. This is by construction because we specifically chose
non-BZ-like conditions for these simulations -- non-spinning BH, weak
magnetic field well below the MAD limit.

The transfer of energy from radiation to gas causes the gas velocity
to accelerate along the funnel, as seen in the bottom panel of
Fig.~\ref{f.enfluxes}. The gas at the axis reaches roughly $50\%$ of
the speed of light at distance $r\sim 1000$. The density-weighted gas
velocity in the funnel reaches only $\sim 35\%$ of the speed of light,
reflecting the fact that gas near the edge of the funnel is denser and
less effectively accelerated.

Acceleration continues as long as momentum is transferred from the
radiation, i.e., up to radius $r\approx 2000$. Constant gas velocity implies that the kinetic energy has
saturated. However, the fractional
contribution of the kinetic energy (purple line in the top panel)
increases even further. This is because of the residual radiation crossing the borders of the
jet region and increasing the kinetic contribution. Acceleration stops when radiation drag by photons from
the funnel walls is large enough to counter-balance the acceleration
by radiation flowing along the funnel \citep{sikorawilson-81}. In the
framework of the M1 closure scheme, this corresponds to the situation
where the gas comoves with the radiation rest frame, where no momentum
is transferred between the radiation and the gas. 

\begin{figure}
 \includegraphics[width=.97\columnwidth]{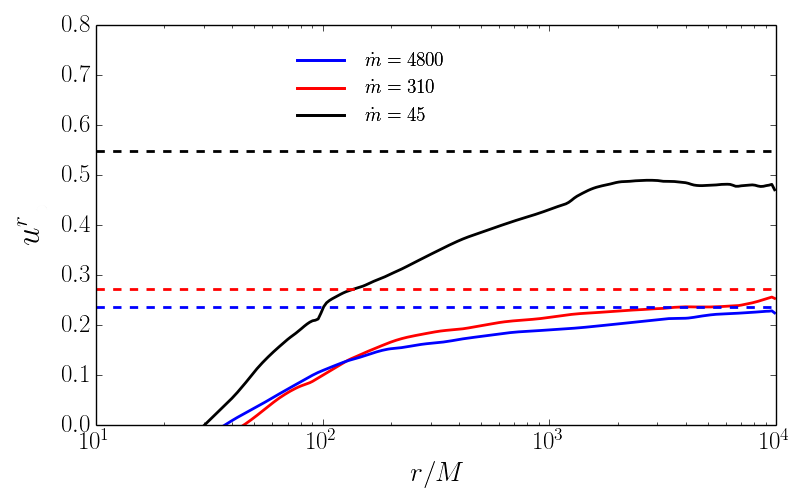}
  \caption{ Radial velocity of the gas at the axis (solid lines) and
    the maximal velocity the gas will achieve if all the energy is
    converted into kinetic energy at infinity (dashed lines). Results
    are shown for all three models.  }
 \label{f.Bevel}
\end{figure}

Even though we ran our simulations over a large domain extending to
$r_{\rm out}=10000$, it turned out that the outer boundary was still
not far enough out to resolve the photospheres of Models \texttt{B}
and \texttt{C}. In these models, gas acceleration occurs all the way
to the boundary. Nevertheless, we can still estimate the terminal
velocity of the gas by looking at the total energy flux near the
boundary and assuming that all the energy is converted into kinetic
energy at infinity. The velocities along the axis and the terminal
velocities for the three models are shown in Fig.~\ref{f.Bevel} by
solid and dashed lines, respectively. The gas in the lowest accretion
rate model \texttt{A} reaches saturated velocity $0.5c$ already at
radius $r=1000$. The terminal velocity for this model (black dashed
line) is only $\sim 10\%$ higher, which means that almost all the
available energy has been converted into kinetic energy. For the other
two models, with accretion rates $310$ and $4800\Medd$, the actual and
terminal gas velocities are significantly lower and reach only $20$ to
$30\%$ of the speed of light. This is probably the result of more
effective mass loading of the funnel region in these models. Because
of the larger optical depth, we presume radiation more effectively
drives gas towards the axis, which results in weaker acceleration of
the gas in this region.

\subsection{Luminosities}
\label{s.luminosities}

\begin{figure}
 \includegraphics[width=1.1\columnwidth]{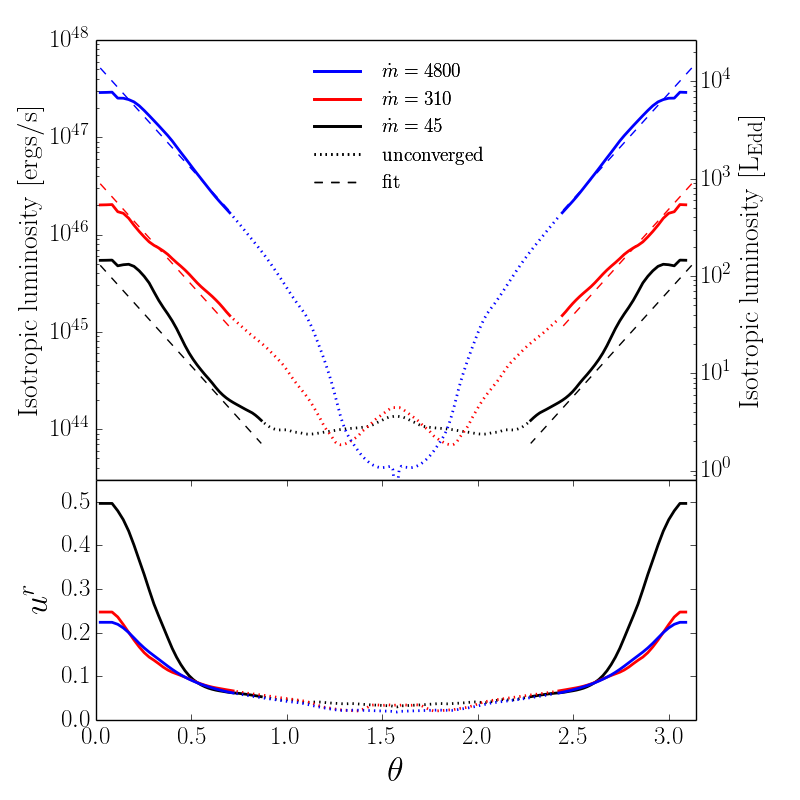}
  \caption{Top panel: Isotropic equivalent luminosity (measured at
    $r=5000$) for models \texttt{A} (black), \texttt{B} (red) and
    \texttt{C} (blue lines) as a function of the polar angle
    $\theta$. The dashed lines show the fitting function given in
    eq.~\ref{e.lumfit} for a black hole mass
    $M=3\times10^5M_\odot$. Bottom panel: Velocity of the gas at
    $r=5000$. In both panels the solid lines correspond to the region
    of the simulation that is in inflow/outflow equilibrium at
    $r=5000$, while the dotted lines refer to unconverged regions.}
 \label{f.lumvsth}
\end{figure}

Fig.~\ref{f.lumvsth} presents the equivalent isotropic luminosity (top
panel, Eq.~\ref{e.isolum}) and the radial velocity of the gas (bottom
panel) as a function of polar angle for models \texttt{A} ($\dot{M} =
45 \dot{M}_{\rm Edd}$), \texttt{B} ($310 \dot{M}_{\rm Edd}$), and
\texttt{C} ($4800\Medd$). The luminosities were measured at radius
$r=5000$ and reflect the flux of total energy, which in the funnel/jet
region is dominated by kinetic luminosity.  The solid lines correspond
to the region of the simulation domain where inflow/outflow
equilibrium state has been reached at $r=5000$. The dotted lines refer
to unconverged region, which is not in flow contact with the BH and
therefore the results are less reliable.

For an observer looking precisely down the axis, $L_{\rm iso} =
5\times 10^{45}\rm erg\,s^{-1}$ for model \texttt{A}, and increases to
as much as $3\times 10^{47}\rm erg\,s^{-1}$ for model \texttt{C}
($4800\Medd$). The isotropic equivalent luminosity is lower for
off-axis observers, dropping by roughly an order of magnitude between
the axis and the border of the jet region. The luminosities scale
linearly with accretion rate, i.e., the efficiency of energy injection
into the funnel region seems to be independent of the accretion rate.

The isotropic equivalent luminosity profiles are reasonably well
described by the following fitting function, \be L_{\rm fit}=4\times 10^{47}
e^{-\theta/\theta_{\rm jet,0}} \frac{\dot M}{10^3\dot M_{\rm Edd}}\frac{M_{\rm
    BH}}{10^6M_\odot},
\label{e.lumfit}
\ee 
where the scaling with BH mass is included\footnote{Although we
  simulated accretion flows on a supermassive BH (to directly show how
  strong the radiative jet can be) our results can be applied to any
  astrophysical BH by scaling them properly with mass. This is because
  the optical depth over given number of gravitational radii, which
  determines, e.g., the efficiency of gas/radiation interaction, is
  independent of BH mass for fixed fraction of Eddington accretion
  rate.} and where $\theta_{\rm jet,0} = 0.2$ is the characteristic
jet beaming angle\footnote{Allowing for the fact that there is more
  solid angle at larger $\theta$, a better estimate of the luminosity
  beaming angle is twice this: $\theta_{\rm jet} \approx 0.4$.}.

The bottom panel of Fig.~\ref{f.lumvsth} shows the dependence of the
gas velocity on polar angle for each of the models. Model \texttt{A}
with the lowest accretion rate shows the most efficient
acceleration. In all three models, the velocity quickly decreases with
increasing $\theta$ and transitions to $u^r < 0.1$ in the wind
region ($\theta>0.5$).

\subsection{Beaming factor}
\label{s.beaming}

Disks accreting at super-critical rates are geometrically very
thick. As a result, energy in general and photons in particular escape
preferentially along the axis, where the density and optical depth are
lowest. Most of the energy escapes through the funnel, and the flux of
energy decreases steeply with increasing polar angle
(Eq.~\ref{e.lumfit}, Fig.~\ref{f.lumvsth}). It is useful to define the
beaming factor $b(\theta)$, which gives the ratio of the isotropic
equivalent luminosity $L_{\rm iso}$ (defined in eq.~\ref{e.isolum})
inferred by an observer at polar angle $\theta$ to the real luminosity
of the source $L_{\rm tot}$,
\be b = \frac{L_{\rm iso}}{L_{\rm tot}}=\frac{L_{\rm iso}}{\eta \dot M
  c^2},
\label{eq.b}
\ee where $L_{\rm tot}$ is obtained by integrating the energy flux
over the sphere\footnote{In steady state, the integrated
    energy flux is independent of radius, so it can be evaluated at
    any convenient radius. The values of $L_{\rm tot}$ and $\eta$
    quoted here correspond to $r = 5000$.}, 
\be 
L_{\rm tot}=\int F^r \sqrt{-g}\,{\rm d}\theta{\rm d}\phi \equiv \eta
\dot M c^2, 
\ee 
with $\eta$ being the total efficiency of the
accretion flow (results given in Table~\ref{t.models}).  When
integrated over the sphere, the beaming factor should, on average, be
equal to unity. However, because the equilibrium region at $r=5000$
  is 
limited to the polar region (gas near the equatorial plane was not
able to relax yet because of long viscous timescale there), this
integral cannot be directly calculated.

\begin{figure}
 \includegraphics[width=1.\columnwidth]{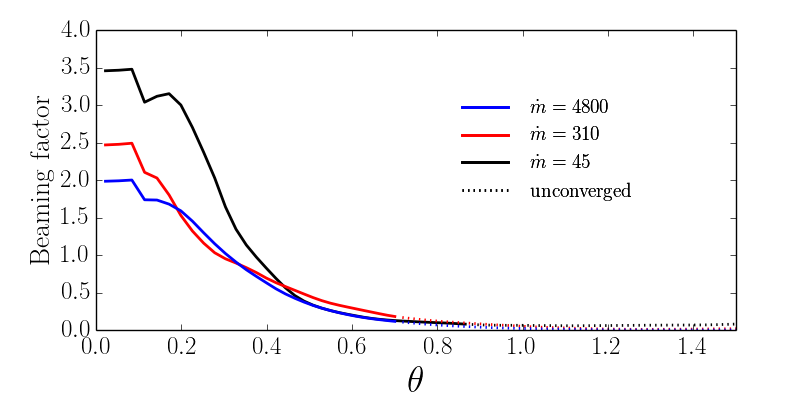}
  \caption{Beaming factor (for definition see Eq.~\ref{eq.b}) as a
    function of polar angle in the three simulations.  }
 \label{f.bvsth}
\end{figure}

Fig.~\ref{f.bvsth} shows the beaming factor as a function of $\theta$,
as measured at radius $r=5000$. For all three runs, $b$ is maximum at
the axis and decreases away from the axis. The maximal beaming factor
is $b=3.5$ for model \texttt{A}, and $b=2.0-2.5$ for models \texttt{B}
and \texttt{C}. More pronounced beaming in model \texttt{A} may come
from the fact that, in this case, $b$ is measured outside the
photosphere. The beaming factor equals unity, i.e., local isotropic
equivalent luminosity is equal to the true luminosity, at $\theta
\approx 0.35$, which roughly coincides with the edge of the funnel. In
the wind and disk regions, $b(\theta) \ll 1$, reflecting the fact that
comparable amounts of energy leave in the funnel and in these regions,
but the solid angle covered by the funnel is a small fraction of the
whole sphere.

It should be emphasized that we defined the beaming factor $b(\theta)$
with respect to the {\it total luminosity} of the accretion flow, not
the Eddington luminosity. The three simulations discussed in this
paper are all energetically efficient, as reflected by the fact that
they have efficiencies $\eta$ (Table~\ref{t.models}) nearly equal to
the standard Novikov-Thorne thin disk efficiency, $\eta_0 =
0.057$. Correspondingly, $L_{\rm tot}/L_{\rm Edd} \approx \dot{M} /
\dot{M}_{\rm Edd}$. Therefore, if we define an Eddington-scaled
beaming factor $b_{\rm Edd}(\theta)$, we roughly expect \be b_{\rm
  Edd} \equiv \frac{L_{\rm iso}}{L_{\rm Edd}} \approx
\frac{\dot{M}}{\dot{M}_{\rm Edd}}\,b.  \ee Considering that $b$ itself
is of order a few along the axis, it is clear that $b_{\rm Edd}$ can
be extremely large for highly super-Eddington accretion flows. If the
BH was spinning and if it was threaded by a MAD-level magnetic field,
the luminosity would be larger still by possibly a factor of several.

\section{Discussion}
\label{s.discussion}

Magnetically-dominated, BH rotation-powered jets in radiatively
inefficient hot accretion flows have been simulated and studied now
for a number of years. These jets operate through the BZ mechanism, in
which ordered magnetic fields are twisted by the rotating BH. The
twisted field lines unwind and expand in a direction parallel to the
BH spin axis because of the collimating action of the geometrically
thick accretion disk.

The jets studied in this paper involve different physics. Our BHs do
not rotate and there is no strong ordered magnetic field, so there is
no opportunity for the BZ mechanism to operate. Nevertheless, the
super-Eddington accretion flows considered here do produce powerful
jets. Because these accretion disks are geometrically thick, they
naturally form funnels oriented parallel to the angular momentum axis
of the accretion flow (just as in the case of radiatively inefficient
hot accretion flows). MRI-driven turbulence in the disk leads to
energy dissipation in the disk interior, which heats the gas. The gas,
in turn, cools by emitting radiation. The radiation then flows down
the gradient of density/optical depth via diffusion and turbulent
transport \citep{jiang+14}, which naturally concentrates a significant
amount of energy in the funnel region. There, radiation pressure
accelerates gas outward along the axis of the funnel, resulting in a
jet that travels at a modest fraction (0.2--0.5) of the speed of
light.

The process described above will take place so long as the accretion
flow forms a thick disk with a funnel. Independently of whether the
funnel is optically thin or thick, radiation always moves towards the
axis. Once there, it collimates and rushes out along the axis,
transfering its momentum to the gas. The transfer operates as long as
there is enough optical depth. Once the radiation crosses the
photosphere, where radiative and kinetic energies are of the same
order (see Fig.~\ref{f.enfluxes}), the acceleration of the gas
terminates. 

Efficient conversion of radiative to kinetic energy happens only if
there is sufficient optical depth along the funnel. Already for an
accretion rate of $45\Medd$ (model \texttt{A}, the lowest $\dot{M}$ we
have considered), most of the radiative energy is converted to kinetic
energy. For lower accretion rates, the optical depth in the funnel
will be lower, and we expect that a larger fraction of the jet energy
will remain in the form of radiation rather than gas kinetic
energy. However, the total jet power should scale similarly for all
super-critical accretion rates. This is because the efficiency of
accretion and the thickness of the disk (which collimates the jet) are
independent of the accretion rate in this regime
\citep{sadowski+dynamo}.

\begin{figure}
 \includegraphics[width=1.\columnwidth]{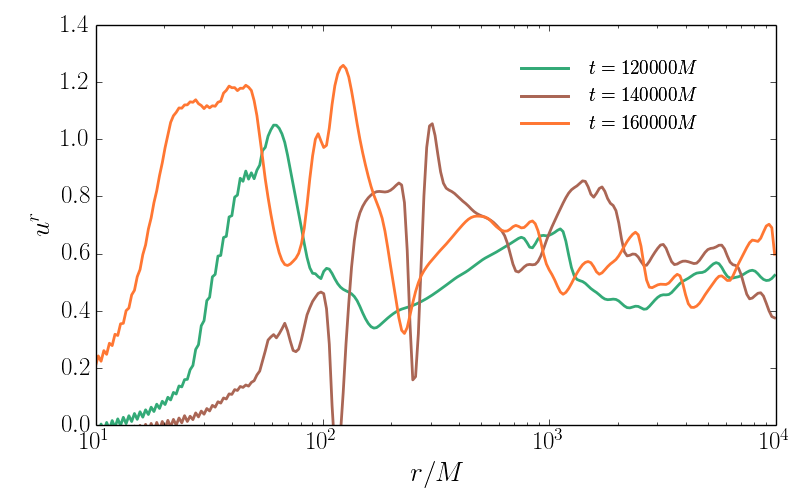}
  \caption{Velocity profiles along the axis for model \texttt{A}
    corresponding to three epochs.  }
 \label{f.velvsz}
\end{figure}

Can the kinetic energy escaping the funnel be observed by a distant
observer? Yes, provided it is converted back into radiation. In
magnetic jets, reconnection is possibly a dominant mechanism for this
conversion \citep{sironi+15}. However, magnetic reconnection is
unlikely to be important for the radiatively-driven jets considered
here since there is so little magnetic energy (Fig.~3). Shocks are a
more natural mechanism for converting the gas kinetic energy to
radiation via nonthermal particle acceleration. To show that shocks
are quite likely, Fig.~\ref{f.velvsz} presents the radial profile of
gas velocity near the axis for model \texttt{A} at three different
moments of time. We see that the velocity profiles along the funnel
are highly variable. Layers (shells) of gas moving with higher than
average velocity will clearly overtake preceding lower-velocity gas,
thereby producing shocks (especially in optically thin regions).  Such
``internal shocks'' will dissipate a good fraction of the kinetic
energy, converting it to thermal and nonthermal energy and ultimately
into radiation. What fraction of the total energy flux measured at
$r=5000$ (Fig.~\ref{f.lumvsth}) reaches the observer in the form of
radiation depends on the geometry and dynamics of this process.
The energy distribution of the reprocessed radiation can be
significantly different from that of the original radiation in the
funnel. Estimating the spectral shape is, however, beyond the scope of
this paper.

\subsection{Comparison with previous works}

Studies of geometrically thick disks corresponding to high accretion
rates were initiated in the late 1970s by authors studying thick
equilibrium torii
\citep{fishbone+76,kozlowski+78,abramowicz+78,jaroszynski+80}. It was
realised that if torii are thick and radiative pressure supported,
their total luminosity must exceed the Eddington limit
\citep{paczynskiwiita-80}.  Later works by \cite{sikora-81},
\cite{sikorawilson-81} and \cite{narayan+83} specifically analysed the
properties of the polar outflows. The accretion flows were modeled by
thick equilibrium torii with arbitrarily prescribed angular momentum
profiles. It was shown that, if such torii are thick enough, the local
radiative flux in the funnel may significantly exceed the Eddington
value. \cite{sikorawilson-81} showed in addition, that the flow cannot
convert more than a small fraction of the disk's radiative luminosity
into kinetic energy in the polar outflow because of radiation
drag. However, the authors noted that the acceleration may be more
efficient if the funnel is optically thick. These predictions are in
perfect agreement with our study.

Numerical simulations of super-critical accretion are relatively new.
The field was initiated in a series of seminal works by
\cite{ohsuga05, ohsuga09, ohsuga07, ohsuga11}, who performed radiation
hydrodynamical, and later MHD, simulations using a non-relativistic
code and adopting the flux-limited diffusion approximation. More
recently, our group developed general relativistic numerical codes
with M1 radiative closure and applied it to super-critical accretion
\citep{sadowski+koral2,sadowski+dynamo,mckinney+harmrad}. A similar
scheme was implemented in an independent radiation GRMHD code by
\cite{fragile+14}.  Most recently, \cite{jiang+14} simulated a
super-critical accretion flow using a Newtonian code with cylindrical
coordinates (and a cylindrical BH). The authors solved the radiative
transfer equation directly on a fixed grid of angles, instead of
applying any radiative closure. 

All the studies cited above agree qualitatively on the main features
of super-critical accretion, viz., that there is no barrier to
super-Eddington accretion, that these flows are optically and
geometrically thick, and that they are characterized by strong polar
outflows and significant winds. The detailed composition of the
outflows differs between studies, e.g., in \citet{jiang+14} energy
flows out primarily via radiation, whereas our jets are kinetic energy
dominated, and the reason for this is presently unclear. Our work is
the first one to study polar outflows using a general relativistic
code. Also, we simulated accretion onto a (mildly) supermassive BH
(all previous work considered stellar-mass BHs).

In a work similar to ours, \cite{takeuchi+10} studied radiatively
driven jets using a Newtonian code and simulated an accretion flow
with $\dot M\approx 6\Medd$. The authors focused on the acceleration
and collimation mechanisms of the radiative jet. They showed that the
jet can be accelerated up to mildly relativistic speeds by the
radiation-pressure force and collimated by the Lorentz force of a
magnetic tower structure in the innermost region. They also pointed
out the importantce of the radiative drag force in limiting gas
acceleration in \cite{takahashiohsuga-15}. The gas velocities they
obtained are in perfect agreement with our study. However, we do not
observe magnetic collimation of the outflow. This discrepancy may be
related to the fact that the simulated flows in the present work have
significantly higher accretion rates.

\subsection{Astrophysical implications}
\label{s.implications}

A principal result of this paper is that super-critical accretion on
BHs leads to a collimated outflow of energy along the polar axis even
if the BH spin is zero and BZ-like extraction of BH rotational energy
is ruled out. For observers looking directly down the funnel, the
isotropic-equivalent luminosity can significantly exceed the Eddington
luminosity (Eq.~\ref{e.lumfit}).  The observed luminosity drops with
increasing polar angle. We discuss the importance of these results in
the context of tidal disruption events (TDEs), ultra-luminous X-ray
sources (ULXs), and the microquasar SS433.

\subsubsection{Tidal disruption events}

Whenever a star comes closer to a supermassive BH than its tidal
radius, it is disrupted by the tidal gravitational field
\citep{rees-88,evanskochanek-89}. Approximately half the mass of the
star remains bound to the BH and returns after a while to the vicinity
of the BH. Relativistic precession causes the tidal stream to interact
with itself, shock and circularize \citep[e.g.,
][]{bonnerot+15,shiokawa+15}. Once the gas circularizes, it starts to
build up magnetic field which triggers accretion. According to the
standard model \citep[e.g.,][]{krolikpiran-12}, for typical parameters
(stellar mass star disrupted by a $\sim10^6 M_\odot$ BH), the peak
accretion rate at the onset of accretion can exceed $10^3\Medd$. This
accretion rate subsequently declines as $t^{-5/3}$. 

In the case of the TDE Swift J1644, the observed isotropic-equivalent
X-ray luminosity was as high as $10^{48}\rm erg\,s^{-1}$ during the
initial outburst peak \citep{bloom11,burrows11}.  This fact, together
with the properties of the radio-emission, led to the suggestion that
the extreme luminosity was produced by a relativistic jet driven by
rotational energy of the BH. However, efficient extraction of BH spin
energy requires the accumulation of significant magnetic flux at the
BH, far more than the disrupted star can provide
\citep{tchekh+14}. This problem could be overcome if the tidal stream
is able to drag magnetic field from a pre-existing accretion flow
around the BH \citep{kelley+14}.

The initial highly super-Eddington regime of accretion in a TDE
corresponds precisely to the simulations discussed in this study,
where we consider super-critical accretion on a supermassive BH with
$\dot{M}$ of order tens to thousands of Eddington. We can therefore
explore whether a purely radiative jet, similar to the ones studied in
this work, can explain observations of TDE jets. As shown in
Section~\ref{s.luminosities}, for an observer looking down the funnel,
the isotropic equivalent luminosity reaches (Eq.~\ref{e.lumfit})
$10^{47}\,\rm erg\,s^{-1}$ for canonical TDE parameters. By increasing
the BH mass by an order of magnitude to $10^7M_\odot$, it is possible
to match the peak luminosity observed in Swift J1644. However, the
radio emission in the afterglow of this TDE event suggests that the
jet must have been relativistic, with Lorentz factor $\gtrsim 2$
\citep{berger+12}. Such high velocities are difficult with purely
radiative acceleration --- the radiation field in the funnel is not
sufficiently collimated, and acceleration is prevented by radiative
drag. Therefore, it seems that radiative jets alone cannot explain the
properties of Swift J1644. Instead, the jet in this object probably
involved a combination of radiatively and BH rotational energy driven
acceleration.

Some other TDE candidate events, e.g., PS1-10jh \citep{gezari+12},
PS1-11af \citep{chornock+14}, and PTF09ge \citep{arcavi+14}, did not
show significant X-ray and radio emission, but were observed in the
ultraviolet band only. This fact may be explained if we assume that
these sources do not produce strong magnetic jets accelerating gas to
such high velocities as in the case of Swift J1644.  Having in mind
that the onset of accretion of the returning debris corresponds to
strongly super-critical accretion, and that the luminosities are
super-Eddington, the observed optical emission would come from the
funnel region where it is reprocessed through the kinetic flux, as
discussed in this work. It could also come directly from the inner
funnel wall provided the optical depth in the polar region is low,
though this is likely only at lower accretion rates. In both cases, it
is required that the TDE candidates must be observed within the jet
beaming angle, because the luminosity would be significantly lower for
more inclined observers (Fig.~\ref{f.bvsth}).

\subsubsection{Ultraluminous X-ray sources}

Ultraluminous X-ray sources (ULXs) are objects emitting isotropic
equivalent luminosities in excess of $10^{39}\,\rm erg/s$
\citep[e.g.,][]{miller+04}. The critical number corresponds roughly to
the Eddington luminosity for a $10 M_\odot$ central source.  A
standard thin disk around a $10M_\odot$ BH, accreting at a
sub-Eddington rate, cannot produce the luminosity observed in ULXs, so
one possibility is that the accreting compact objects are
intermediate-mass BHs. However, if the emitted radiation is beamed
into a relatively narrow solid angle, the observed luminosity can
easily exceed the Eddington value for observers located in favorable
directions, even though the total emission may still be sub-critical.
In this scenario, ULXs are normal $\sim10M_\odot$ BHs \citep{king-09}.

Our study supports the latter picture since we find very strong
beaming of radiation in geometrically thick disks. The particular
systems we studied have very high and extreme accretion rates. In
these flows the funnel region is optically thick and radiative energy
is efficiently converted into kinetic energy of the outflowing,
collimated gas. Whether or not this kinetic energy can be
dissipated in shocks to produce radiation and whether the resulting
spectrum would match observations of ULXs is uncertain.

However, another regime of super-critical accretion may be more
applicable to ULXs. For not-so-extreme accretion rates, say
$\dot{M}\sim10 \Medd$, the polar region is optically thin and
conversion of radiative to kinetic energy is not efficient. It is then
possible to observe directly the radiation emitted inside the
funnel. Although the beaming may be less strong, it can still magnify
the observed luminosity and cause a stellar-mass BH source accreting
at a moderately super-critical rate to match the isotropic-equivalent
luminosities observed in ULXs.  Detailed modeling of the observational
properties of such systems was done by \cite{kawashima+12} using
simulation data corresponding to a mildly super-critical ($\sim 10
\Medd$) accretion flow on a $10M_\odot$ BH. The simulation was done
with a Newtonian radiation MHD code. The authors showed that
Comptonization in the polar region significantly hardens the original
photon spectrum.

We have estimated the beaming factor, defined as the ratio of the
observed (isotropic equivalent) to the real (integrated over the
sphere) luminosities, and find (Fig.~\ref{f.bvsth}) that it
significantly decreases with increasing polar angle. Therefore, one
should expect that, for the same BH mass and accretion rate, the
distribution of ULX luminosities will be sensitive to the orientation
on the sky.

\subsubsection{SS433}

SS433 is the first microquasar to be discovered. It is an eclipsing
X-ray binary system in which the compact primary object is most likely
a black hole with $M_{\rm BH}\approx 10M_\odot$
\citep{grindlay+84}. The wind of the companion star feeds the central
object at an extremely high rate.  The measured disk wind outflow
rates are $\dot M=10^{-5} - 10^{-4}\, \dot M_\odot {\rm yr^{-1}}$
\citep{shklovsky-81} which suggests that the accretion rate on to the
compact object is super-critical. A direct measurement of the
accretion $\dot{M}$ is, however, impossible because the innermost
regions of the disk are obscured.  SS433 shows powerful jets
precessing with a period of 162 days and sweeping twin cones with
half-angle $20^\circ$. The jets are loaded with baryons and move at a
velocity of $0.26c$ \citep{margon+84,fabrika-04}. The corresponding
kinetic power dominates the jet energy budget and equals $10^{39.3} -
10^{39.7}\,\rm erg\,s^{-1}$ \citep{marshall+15}.

The fact that the jets are precessing suggests that the BH spin is
non-zero and that there is a misalignment between the angular momentum
of the binary and the BH spin axis. These conditions have to be
satisfied for Lense-Thirring precession \citep{bardeenpetterson-75} to
occur. If the BH is spinning, then it seems reasonable to assume that
the observed jets are powered by the BH rotational energy via a strong
magnetic field. However, the origin of the baryons in the jet is then
problem since the magnetic field lines will screen the baryons in the
disk from the jet region. It has been suggested that the jets are
initially leptonic, but they then acquire baryons when they collide
either with the thick accretion disk itself or the disk wind.

We propose another explanation for the baryon-loaded jets in
SS433. The mass supply rate on the compact object most likely
corresponds to a super-critical accretion. At such high accretion
rates, an optically thick and geometrically thick disk will form and,
as shown in this work, radiation moves into the polar region, pushing
gas along with it and making the funnel optically thick. The radiation
then escapes up the funnel and accelerates the gas, thus converting a
good fraction of the total energy flux into jet kinetic energy. Once
such a radiatively-driven jet crosses the funnel photosphere, it is
baryon-loaded, kinetically-dominated and collimated.

These features in our simulations are consistent with observations of
the jets in SS433. The density-averaged gas velocity in our simulated
jets is of the order of $0.2-0.3c$ (bottom panel of
Fig.~\ref{f.enfluxes}), which is in excellent agreement with the
velocity observed in SS433.  For a $10M_\odot$ BH accreting at
$100\Medd$, the jet power we expect based on the simulations is
roughly $0.01\dot Mc^2 \approx 2\times 10^{40}\, \rm erg\,s^{-1}$.
This is more than adequate to explain the observed kinetic luminosity
of the jets in SS433.

The non-zero BH spin and misalignment between the orbital and BH spin
axes, both of which are required for jet precessions, suggest that
additional effects over and above simple radiative acceleration might
be operating.
The small observed opening angle of the jet in SS433 ($1^\circ-
2^\circ$), is not consistent with the jets seen in the simulations,
which have opening angles of the order of $10^\circ - 15^\circ$
(Fig.~\ref{f.bvsth}). Perhaps there are magnetic effects on top of the
radiative processes we have discussed, which cause strong collimation.

\section{Acknowledgements}

The authors thank James Guillochon, Herman Marshall, Tsvi Piran, Nir Shaviv and Rashid
Sunyaev for helpful discussions.  AS acknowledges support for this
work from NASA through Einstein Postdoctotral Fellowship number
PF4-150126 awarded by the Chandra X-ray Center, which is operated by
the Smithsonian Astrophysical Observatory for NASA under contract
NAS8-03060. AS thanks Harvard-Smithsonian Center for Astrophysics for
hospitality.  RN was supported in part by NSF grant AST1312651 and
NASA grant TCAN NNX14AB47G.  The authors acknowledge computational
support from NSF via XSEDE resources (grant TG-AST080026N), and from
NASA via the High-End Computing (HEC) Program through the NASA
Advanced Supercomputing (NAS) Division at Ames Research Center.
 
\bibliographystyle{mn2e}

\end{document}